\documentclass{article}
\usepackage{spconf,amsmath,graphicx}
\usepackage{bm}
\usepackage{subfigure}
\usepackage{float}
\usepackage{multirow}
\usepackage{booktabs}

\title{Squeezing value of cross-domain labels: a decoupled scoring approach for speaker verification}
\name{Lantian Li, Yang Zhang, Jiawen Kang, Thomas Fang Zheng, Dong Wang
}
\address{Center for Speech and Language Technologies, Tsinghua University}
\begin{document}
%
\maketitle
\begin{abstract}

Domain mismatch often occurs in real applications and causes serious performance reduction on speaker verification systems.
The common wisdom is to collect cross-domain data and train a multi-domain PLDA model, with the hope to learn 
a domain-independent speaker subspace. 
In this paper, we firstly present an empirical study to show that simply adding cross-domain data does not help 
performance in conditions with enrollment-test mismatch. Careful analysis shows that this striking result is caused by 
the incoherent statistics between the enrollment and test conditions. Based on this analysis, we present a decoupled scoring 
approach that can maximally squeeze the value of cross-domain labels and obtain optimal verification scores 
when the enrollment and test are mismatched. 
When the statistics are coherent, the new formulation falls back to the conventional PLDA. 
Experimental results on cross-channel test show that the proposed approach
is highly effective and is a principle solution to domain mismatch.

\end{abstract}
\section{Introduction}
\label{sec:intro}

Speaker verification aims to recognize claimed identities of speakers.
After decades of research, current speaker verification systems have achieved
rather satisfactory performance, especially 
with the i-vector model~\cite{dehak2011front}, or embedding models based on 
deep neural nets (DNNs), e.g., the x-vector model~\cite{snyder2018xvector,okabe2018attentive}.
The deep embedding models have been significantly improved recently,
by employing advanced techniques such as novel architectures~\cite{chung2018voxceleb2,Jung2019raw},
attentive pooling~\cite{okabe2018attentive,Xie19a,Chen2019tied},
or max-margin training~\cite{ding2018mtgan,Wang2019centroid,bai2019partial,Gao2019improving}.
As a result, deep learning models have achieved state-of-the-art performance on several benchmark datasets~\cite{Sadjadi2019},
in particular when combined with the PLDA model~\cite{Ioffe06} for scoring.

In spite of the great achievement, a large performance degradation is often observed when current speaker verification systems
are deployed to real applications.
A particular problem is domain mismatch, including the training-deployment domain mismatch and enrollment-test domain mismatch.
For example, when the enrollment uses one recording device and the test uses another device, the performance will be seriously 
degraded.


A large body of research has been conducted to solve domain mismatch problem,
and the basic idea among these approaches is either normalization or adaptation:
the former aims to make the data or model domain-independent
~\cite{rahman2018improving,wang2018unsupervised,Williams2019ff,bhattacharya2019adapting,kang2020disentangled}, 
while the latter aims to make them more suitable for the target domain
~\cite{shon2017autoencoder,zhang2018analysis,lee2019coral}.
Among all these approaches, training a multi-domain PLDA is commonly used. This approach 
collects data from multiple domains and then trains the PLDA model, with the hope to learn 
a domain-independent speaker subspace. 
This PLDA multi-domain training (MDT) is in particular useful when the training data 
is too limited to conduct more data-hungry approaches, e.g., the domain-invariant feature learning~\cite{wang2018unsupervised,kang2020disentangled}.
A key advantage of this MDT approach is that the resultant PLDA model will be suitable for both training-deployment mismatch and enrollment-test mismatch.
Besides, domain-adaption training (DAT) is another approach, which transforms speaker vectors from source domain to target domain, 
and then performs the scoring process in target domain. 
The transform can be established based on maximum likelihood estimation.


For both MDT and DAT, cross-domain speakers are essentially important. The cross-domain labels provide information 
regarding the variation related to domains. A common wisdom is to collect as much cross-domain data as possible, 
with the hope to learn a \emph{true} speaker subspace that is independent of domain change. However, our experimental 
study shows that this is not true: more cross-domain data does not necessarily lead to better performance in conditions 
with enrollment-test mismatch. We give a careful analysis on this phenomenon, and find that it is related to the 
statistics incoherence problem caused by domain mismatch. Based on this analysis, 
we propose a decoupled scoring approach. The key idea is to decouple the scoring process into different phases
and utilize its own correct statistics in each phase. 
By this model, the cross-domain labels are only used to establish the link between the \emph{statistics of different domains}. 


The rest of this paper is organized as follows. Section~\ref{sec:exp} presents experimental setups,
and Section~\ref{sec:analy} gives an empirical study on the statistics incoherence problem.
Section~\ref{sec:decop} presets details of our proposed decoupled scoring approach.
Section~\ref{sec:result} describes experimental results, and the entire paper is concluded in Section~\ref{sec:con}.

\section{Experimental setups}
\label{sec:exp}

\subsection{Data}

Two datasets were used in our experiments: VoxCeleb~\cite{chung2018voxceleb2,nagrani2017voxceleb} and
AIShell-1~\cite{aishell_2017}.
VoxCeleb was used to train the speaker embedding model, which is the x-vector model in our experiment.
AIShell-1 was used for performance evaluation under cross-channel domain mismatch scenarios.
More information about datasets is presented below.

\emph{VoxCeleb}:
The entire dataset contains $2,000$+ hours of speech signals from $7,000$+ speakers.
Data augmentation was applied to improve robustness, with the MUSAN corpus used to generate noisy utterances,
and the room impulse responses (RIRS) corpus was used to generate reverberant utterances.

\emph{AIShell-1}: 
This is an open-source multi-channel Chinese Mandarin speech dataset published by AISHELL.
All the speech utterances are recorded in parallel via three categories of devices,
including high fidelity Microphones (Mic), Android phones (AND) and Apple iPhones (iOS).
This dataset is used for the cross-channel (domain) test in our experiment.
The entire dataset consists of two parts: \emph{AIShell-1.Train}, which covers $3$ devices and involves $360,897$ utterances from $340$ speakers,
was used to implement both the ad-hoc normalization/adaptation methods and our proposed decoupled scoring method.
\emph{AIShell-1.Eval}, which also covers $3$ devices and involves $64,495$ utterances from $60$ speakers,
was used for performance evaluation on cross-channel conditions.

\subsection{Embedding models}

We built the state-of-the-art x-vector embedding model based on the TDNN architecture~\cite{snyder2018xvector}.
This x-vector model was created using the Kaldi toolkit, following the VoxCeleb recipe.
The main architecture contains three components, 
including feature-learning component, statistical pooling component and speaker-classification component.
Once trained, the $512$-dimensional activations of the penultimate layer are read out as an x-vector.

\section{Empirical analysis}
\label{sec:analy}

We firstly employ MDT approach to train a domain-independent PLDA.
In detail, data of speakers collected from two domains are labeled in two ways.
One is domain-independent labeling, where data from different domains shares the same speaker label.
The other one is domain-dependent labeling, where data from different domains is relabeled as 
different speakers even if they are from the same speaker.
The PLDA model will be trained in a controlled way, where we control the proportion of 
domain-independent and domain-dependent labels. 
The expectation is that more domain-independent labels will lead to better performance at least in cross-domain test. 

In our experiments, MDT is conducted using \emph{AIShell-1.Dev}, and the resultant PLDA is evaluated on \emph{AIShell-1.Eval}.
Results are shown in Table~\ref{tab:plda}. Note that in the first column, the case name `A-B' means enrollment on condition A and test 
on condition B. The `Base' column presents results using PLDA trained with the enrollment-matched development dataset. 
A key observation is that the best performance is obtained when the PLDA model is 
trained with 40\%-60\% domain-independent labels plus 60\%-40\% domain-dependent labels,
rather with 100\% domain-independent labels. 

\vspace{-2mm}
\begin{table}[htb!]
  \caption{EER(\%) results with MDT under different cross-channel labels.}
  \label{tab:plda}
  \centering
  \scalebox{0.78}{
  \begin{tabular}{llcccccc}
    \cmidrule(r){1-8}
    \multirow{2}{*}{Cases}  & \multirow{2}{*}{Base} & \multicolumn{6}{c}{Proportion of domain-independent labels} \\
    \cmidrule(r){3-8}
                  &         &  0\%     &   20\%   &  40\%    &  60\%   &  80\%   &  100\%    \\
    \cmidrule(r){1-1} \cmidrule(r){2-2} \cmidrule(r){3-8}
         AND-AND  &  0.797  &  -       &   -      &   -      &   -     &   -     &   -       \\
         AND-Mic  &  2.146  &  3.329   &   1.410  &   1.273  &   \textbf{1.066} &   1.259 &   1.151   \\
         AND-iOS  &  1.425  &  1.642   &   1.104  &  \textbf{0.930}  &   1.029 &   1.170 &   1.161   \\
    \cmidrule(r){1-1} \cmidrule(r){2-2} \cmidrule(r){3-8}
         Mic-AND  &  2.175  &  3.675   &   1.953  &   1.746  &   1.184 &   1.307 &   \textbf{1.161}   \\
         Mic-Mic  &  0.778  &  -       &   -      &   -      &   -     &   -     &   -       \\
         Mic-iOS  &  2.251  &  3.732   &   1.883  &   1.675  &   \textbf{1.255} &   1.349 &   1.293   \\
    \cmidrule(r){1-1} \cmidrule(r){2-2} \cmidrule(r){3-8}
         iOS-AND  &  1.599  &  2.024   &   1.472  &   1.241  &   \textbf{1.156} &   1.274 &   \textbf{1.156}   \\
         iOS-Mic  &  2.216  &  3.697   &   1.651  &   1.476  &   \textbf{1.061} &   1.236 &   1.137   \\
         iOS-iOS  &  0.920  &  -       &   -      &   -      &   -     &   -     &   -       \\
    \cmidrule(r){1-8}
  \end{tabular}}
\end{table}
\vspace{-1.5mm}

This somewhat striking result can be intuitively explained by the statistical nature of the PLDA model. Basically, 
PLDA conducts inference based on the between-class and within-class distributions that are learned from data, 
and if the distributions match the data perfectly, PLDA will derive optimal scores in terms of Bayes risk. 
Unfortunately, MDT learns distributions of the pooled data, which are not ideal for any domain. 
For instance, the within-class variance tends to be large by MDT even if it is small in each of the single domain,
which may lead to unreliable scores. 
This analysis is just intuitive; more theoretical explanation will present in Section~\ref{sec:decop}.

\section{Decoupled scoring}
\label{sec:decop}

We present a theoretical analysis and solution for domain mismatch problem as mentioned in the previous section.
This analysis and solution are based on the normalized likelihood (NL) scoring framework that we proposed recently~\cite{wang2020remark}.

\subsection{Revisit NL scoring}

The task of speaker verification is to test the following two hypotheses regarding to a speaker vector $\pmb{x}$
and check which one is more probable: $\{$$H_0$: $\pmb{x}$ belongs to class $k$; $H_1$: $\pmb{x}$ belongs to any class other than $k$ $\}$.
We therefore define the normalized likelihood (NL) as:

\begin{equation}
\label{eq:nl}
\emph{NL} (\pmb{x}|k)= \frac{p(\pmb{x}|H_0)}{p(\pmb{x}|H_1)} = \frac{p_k(\pmb{x})}{p(\pmb{x})}.
\end{equation}
\noindent where $p(\pmb{x}|H_0)$ is the likelihood of $\pmb{x}$ generated by class $k$, denoted by $p_k(\pmb{x})$.
The posterior for $H_0$ is essentially the likelihood $p_k(\pmb{x})$ normalized by the evidence $p(\pmb{x})$.

Assume a simple linear Gaussian to implement the NL score as follows:

\vspace{-1.5mm}
\begin{equation}
\label{eq:pmu}
p(\pmb{\mu}) = N(\pmb{\mu}; \pmb{0}, \epsilon \mathbf{I})
\end{equation}

\vspace{-1.5mm}
\begin{equation}
\label{eq:px-mu}
p(\pmb{x}|\pmb{\mu}) = N(\pmb{x}; \pmb{\mu}, \sigma \mathbf{I}),
\end{equation}

\noindent where $\pmb{\mu}$ and $\pmb{x}$ represent the speaker mean and speaker vector, respectively;
$\epsilon$ and $\sigma$ are the between-speaker variance and the within-speaker variance, respectively.

With this model, it is easy to derive the marginal probability $p(\pmb{x})$ and the posterior probability $p(\pmb{\mu}|\pmb{x})$ based on the Bayes rule~\cite{wang2020remark}. 
Similarly, the likelihood $p_k(\pmb{x})$ can be computed by marginalizing over $\pmb{\mu}_k$, 
where $\pmb{\mu}_k$ can be estimated from the enrollment samples $\{\pmb{x}^k_1,...,\pmb{x}^k_{n}\}$:

\vspace{-1.5mm}
\begin{eqnarray}
\label{eq:test}
p_k(\pmb{x}) &=&p(\pmb{x}|\pmb{x}^k_1,...,\pmb{x}^k_{n}) \nonumber \\
             &=& \int p(\pmb{x}| \pmb{\mu}_k) p(\pmb{\mu}_k|\pmb{x}^k_1,...,\pmb{x}^k_{n}) \rm{d} \pmb{\mu}_k \nonumber \\
             &=& N(\pmb{x}; \frac{n_k\epsilon}{n_k\epsilon + \sigma} \bar{\pmb{x}}_k, (\sigma + \frac{\epsilon \sigma}{n_k\epsilon + \sigma})\mathbf{I}).
\end{eqnarray}

Finally, a logarithmic NL form can be computed as:

\vspace{-1.5mm}
\begin{eqnarray}
\label{eq:nls}
\log \emph{NL} (\pmb{x}|k) &=& \log p_k(\pmb{x}) - \log p(\pmb{x}) \label{eq:nl-unknown-log} \\
                           &\propto& -\frac{1}{\sigma + \frac{\epsilon \sigma}{n_k \epsilon + \sigma}} ||\pmb{x} - \tilde{\pmb{\mu}}_k||^2 + \frac{1}{\epsilon + \sigma} ||\pmb{x}||^2. \nonumber
\end{eqnarray}

\subsection{Three-phase perspective}

A simple arrangement on the NL score shows that:

\vspace{-1.5mm}
\begin{equation}
\emph{NL} = \frac{p(\pmb{x}|\pmb{x}_1, ..., \pmb{x}_n)}{p(\pmb{x})}=\frac{\int p(\pmb{x}|\pmb{\mu}) p(\pmb{\mu}|\pmb{x}_1, ..., \pmb{x}_n) \rm{d} \pmb{\mu}}{ p(\pmb{x}) }.
\end{equation}

\noindent By this NL form, the score is computed based on three phases:

\begin{itemize}

\item The enrollment phase $p(\pmb{\mu}|\pmb{x}_1, ..., \pmb{x}_{n})$ that produces the posterior of the class mean $\pmb{\mu}$.

\item The prediction phase $p(\pmb{x}|\pmb{\mu})$ that computes the probability of a test sample belonging to a class represented by the class mean $\pmb{\mu}$.

\item The normalization phase $p(\pmb{x})$ that computes the probability that $\pmb{x}$ is produced by all potential classes.

\end{itemize}

Usually, these three phases are based on the same statistical model, under the assumption that the statistics of enrollment and test conditions are the same. 
In this case, the NL score is mathematically equal to PLDA~\cite{wang2020remark}.
In conditions where the enrollment and test are in different domains, the statistics of enrollment and test conditions are different,
and so the three phases should use different statistics, in order to obtain an optimal score. 
This phenomenon is called \emph{statistics coherence} (between scoring phases).

The concept of three-phase scoring provides a clear explanation of the strange behavior we observed in Section~\ref{sec:decop}. 
Firstly notice that although domain-independent 
labels help learn a better between-class distribution, it also destroys the within-class distribution, by making it unnecessarily large. 
In contrast, domain-dependent labels help learn the within-class distribution, but is less useful to
learn the between-class distribution.
Moreover, note that a domain-robust between-class distribution is only requested in the enrollment phase\footnote{ 
For the prediction phase, there is no between-class distribution required; and for the normalization phase, 
we need a between-class distribution that \emph{matches} the within-class distribution in order to represent the marginal distribution of the test data.}.
However, a reasonable within-class distribution is requested for all the three phases. 
Therefore, with a limited amount of data, we need balancing the contribution to the between-class and within-class distributions, 
and so have to reserve a proportion of domain-dependent labels.

\subsection{Decoupled scoring}

Following the three-phase perspective, a principle solution to domain mismatch is to use the respective statistical 
model for each phase, which we call \textbf{domain statistics decomposition (DSD)}.
Firstly, for the enrollment phase, the statistical model of the enrollment condition $\{\epsilon \mathbf{I}, \sigma \mathbf{I}\}$ should be used.
Secondly, for the normalization phase, the statistical model of the test condition $\{\hat{\epsilon} \mathbf{I}, \hat{\sigma} \mathbf{I}\}$ should be used.
Finally, for the prediction phase, a transform is applied to connect the conditional probability $p(\pmb{x}|\pmb{\mu})$
and the posterior $p(\pmb{\mu}|\pmb{x}_1, ..., \pmb{x}_n)$.
For simplicity, we assume this transform is linear:

\vspace{-1.5mm}
\begin{equation}
\label{eq:linear}
\pmb{x} = \mathbf{M} \hat{\pmb{x}} + \pmb{b},
\end{equation}
\noindent where $\hat{\pmb{x}}$ represents the observation in the test condition, and $\pmb{x}$ is the transformed data in the enrollment condition.
Obviously, if the transformed data can be well represented by the statistical model of the enrollment condition,
the optimal NL score can be derived. 



According to Eq.(\ref{eq:test}) and Eq.(\ref{eq:linear}), 
we can obtain the prediction probability based on the model of the enrollment condition:

\vspace{-1.5mm}
\begin{eqnarray}
\label{eq:pkx-test}
p_k(\hat{\pmb{x}}; \mathbf{M}, \pmb{b}) &=&p(\mathbf{M} \hat{\pmb{x}} + \pmb{b}|\pmb{x}^k_1,...,\pmb{x}^k_{n}) \nonumber \\
             &=& \int p(\mathbf{M} \hat{\pmb{x}} + \pmb{b}|\pmb{\mu}_k) p(\pmb{\mu}_k|\pmb{x}^k_1,...,\pmb{x}^k_{n}) \rm{d} \pmb{\mu}_k \nonumber \\
             &=& N(\mathbf{M} \hat{\pmb{x}} + \pmb{b}; \frac{n_k\epsilon}{n_k\epsilon + \sigma} \bar{\pmb{x}}_k, (\sigma + \frac{\epsilon \sigma}{n_k\epsilon + \sigma})\mathbf{I}). \nonumber \\
\end{eqnarray}



According to Eq.(\ref{eq:nls}), the NL score is then computed as follows:

\vspace{-1.5mm}
\begin{equation}
\label{eq:nl-gf}
\log \emph{NL} (\hat{\pmb{x}}|k) \propto -\frac{1}{\sigma + \frac{\epsilon \sigma}{n_k \epsilon + \sigma}} ||\mathbf{M} \hat{\pmb{x}} + \pmb{b} - \tilde{\pmb{\mu}}_k||^2 + \frac{1}{\hat{\epsilon} + \hat{\sigma}} ||\hat{\pmb{x}}||^2.
\end{equation}


The optimal parameters $\{\mathbf{M}, \pmb{b}\}$ can be estimated by maximum likelihood (ML) training,
The objective function for this optimization can be written by:

\vspace{-1mm}
\begin{equation}
\label{eq:mle-loss}
\mathcal{L}(\mathbf{M}, \pmb{b}) = \sum_{k=1}^{K} \sum_{i=1}^{N} p_k(\hat{\pmb{x}}_i; \mathbf{M}, \pmb{b}),
\end{equation}
\noindent where $K$ denotes the number of speakers, and $N$ denotes the number of test samples in each speaker.
In our experiment, the Adam optimizer~\cite{kingma2014adam} was used to optimize $\{\mathbf{M}, \pmb{b}\}$.

\section{Experimental results}
\label{sec:result}

In our experiments, the conventional MDT/DAT and our proposed DSD methods are implemented to deal with domain mismatch problem.

\subsection{Basic results}

The basic result with three methods are reported in Table~\ref{tab:channel-test}.
It can be observed that the proposed DSD method consistently outperforms MDT and DAT,
demonstrating that DSD is more effective in dealing with domain mismatch.
The comparison between DSD and DAT is especially interesting, as the two methods look very
similar and only differ in the normalization term $p({\pmb{x}})$.
The clear advantage of DSD demonstrated the solid theory of this decoupled scoring approach.

\begin{table}[htb!]
  \caption{EER(\%) results with three methods on the cross-channel test.}
  \label{tab:channel-test}
  \centering
  \scalebox{0.90}{
  \begin{tabular}{llccc}
     \cmidrule(r){1-5}
       \multirow{2}{*}{Cases} & \multirow{2}{*}{Base}   &  \multicolumn{3}{c}{Methods} \\
                  \cmidrule(r){3-5}
                  &    &  MDT     &  DAT      &  DSD      \\
      \cmidrule(r){1-1} \cmidrule(r){2-2} \cmidrule(r){3-5}
         AND-AND   &   0.797     &  -       &  -        &  -        \\
         AND-Mic   &   2.146     &  1.151   &  1.245    &  \textbf{0.981}    \\
         AND-iOS   &   1.425     &  1.161   &  1.312    &  \textbf{0.623}    \\
      \cmidrule(r){1-1} \cmidrule(r){2-2} \cmidrule(r){3-5}
         Mic-AND   &   2.175     &  1.161   &  1.189    &  \textbf{0.712}    \\
         Mic-Mic   &   0.778     &  -       &  -        &  -        \\
         Mic-iOS   &   2.251     &  1.293   &  1.481    &  \textbf{0.812}    \\
      \cmidrule(r){1-1} \cmidrule(r){2-2} \cmidrule(r){3-5}
         iOS-AND   &   1.599     &  1.156   &  1.184    &  \textbf{0.755}    \\
         iOS-Mic   &   2.216     &  1.137   &  1.231    &  \textbf{1.052}    \\
         iOS-iOS   &   0.920     &  -       &  -        &  -        \\
    \cmidrule(r){1-5}
  \end{tabular}}
\end{table}
\vspace{-1.5mm}

\subsection{Further analysis}

To better understand the capability among three methods to squeeze the value of cross-domain labels, 
different number of speakers are sampled from \emph{AIShell-1.Dev} to train MDT, DAT and DSD respectively.
Results are reported in Table~\ref{tab:comp}.


It can be observed that with the increasing of speaker numbers,
the performance of DSD is consistently improved and gradually outperforms both MDT and DAT.
We attribute its success to the decoupled scoring scheme.
In this scheme, $p_k(\pmb{x})$ and $p(\pmb{x})$ are individually computed.
The $p_k(\pmb{x})$ is optimized based on the statistics in the enrollment condition,
while the $p(\pmb{x})$ is estimated based on the statistics in the test condition.
Therefore, the decoupled NL score has better capability to squeeze the cross-domain data.
In a word, compared to MDT and DAT, DSD can obtain an optimal NL score, and can be regarded as a principle solution to domain mismatch.

\vspace{-1mm}
\begin{table}[htb!]
  \caption{EER(\%) results with three methods under different numbers of cross-domain data.}
  \label{tab:comp}
  \centering
  \scalebox{0.86}{
  \begin{tabular}{clccccc}
    \cmidrule(r){1-7}
    \multirow{2}{*}{Methods} & \multirow{2}{*}{{Cases}} & \multicolumn{5}{c}{\# of speakers} \\
    \cmidrule(r){3-7}
              &           &  68     &  136     &  204     &  272     &  340     \\
    \cmidrule(r){1-1} \cmidrule(r){2-2} \cmidrule(r){3-7}
              &  AND-Mic  &  5.093  &  1.896   &  1.264   &  1.283   &  1.151   \\
              &  AND-iOS  &  5.185  &  1.628   &  1.250   &  1.250   &  1.161   \\
    \cmidrule(r){2-2} \cmidrule(r){3-7}
       MDT    &  Mic-AND  &  5.586  &  2.284   &  1.274   &  1.194   &  1.161   \\
              &  Mic-iOS  &  5.732  &  2.213   &  1.491   &  1.420   &  1.293   \\
    \cmidrule(r){2-2} \cmidrule(r){3-7}
              &  iOS-AND  &  5.213  &  1.807   &  1.236   &  1.151   &  1.156   \\
              &  iOS-Mic  &  5.296  &  1.900   &  1.165   &  1.165   &  1.137   \\
    \cmidrule(r){1-1} \cmidrule(r){2-2} \cmidrule(r){3-7}

    \cmidrule(r){1-1} \cmidrule(r){2-2} \cmidrule(r){3-7}
              &  AND-Mic  &  1.174  &  1.137   &  1.245   &  1.287   &  1.245   \\
              &  AND-iOS  &  1.184  &  1.175   &  1.274   &  1.307   &  1.312   \\
    \cmidrule(r){2-2} \cmidrule(r){3-7}
       DAT    &  Mic-AND  &  1.458  &  1.364   &  1.189   &  1.227   &  1.189   \\
              &  Mic-iOS  &  1.548  &  1.434   &  1.420   &  1.496   &  1.481   \\
    \cmidrule(r){2-2} \cmidrule(r){3-7}
              &  iOS-AND  &  1.222  &  1.123   &  1.146   &  1.208   &  1.184   \\
              &  iOS-Mic  &  1.264  &  1.127   &  1.151   &  1.250   &  1.236   \\
    \cmidrule(r){1-1} \cmidrule(r){2-2} \cmidrule(r){3-7}

    \cmidrule(r){1-1} \cmidrule(r){2-2} \cmidrule(r){3-7}
              &  AND-Mic  &  2.976  &  2.108   &  0.863   &  1.080   &  0.981   \\
              &  AND-iOS  &  1.241  &  1.080   &  0.528   &  0.646   &  0.623   \\
    \cmidrule(r){2-2} \cmidrule(r){3-7}
       DSD    &  Mic-AND  &  5.926  &  5.086   &  0.967   &  0.750   &  0.712   \\
              &  Mic-iOS  &  5.577  &  4.902   &  1.062   &  0.830   &  0.812   \\
    \cmidrule(r){2-2} \cmidrule(r){3-7}
              &  iOS-AND  &  1.840  &  1.632   &  0.646   &  0.816   &  0.755   \\
              &  iOS-Mic  &  2.735  &  1.882   &  0.816   &  1.094   &  1.052   \\
    \cmidrule(r){1-7}
  \end{tabular}}
\end{table}
\vspace{-3mm}

%

\section{Conclusions}
\label{sec:con}

This paper investigated the issue of domain mismatch in speaker verification task,
and found that the statistics incoherence was the essential problem associated with this mismatch.
To deal with this problem, we presented a decoupled scoring approach.
Specifically, we decoupled the scoring process to three phases according to
the normalized likelihood (NL) framework, and used the respective statistical model for each phase.
A simple yet effective linear transform was applied to implement this approach.
Experimental results demonstrated that the proposed decoupled scoring approach was highly effective to squeeze the value of cross-domain data
and obtained the best performance compared to other competitive methods.
Future work will extend this approach to many other mismatch scenarios, e.g., dynamic speaker enrollment, multi-genre test.

\newpage
\bibliographystyle{IEEEbib}
\bibliography{refs}

\begin{thebibliography}{10}

\bibitem{dehak2011front}
Najim Dehak, Patrick~J Kenny, R{\'e}da Dehak, Pierre Dumouchel, and Pierre
  Ouellet,
\newblock ``Front-end factor analysis for speaker verification,''
\newblock {\em IEEE TASLP}, vol. 19, no. 4, pp. 788--798, 2011.

\bibitem{snyder2018xvector}
David Snyder, Daniel Garcia-Romero, Gregory Sell, Daniel Povey, and Sanjeev
  Khudanpur,
\newblock ``X-vectors: {R}obust {DNN} embeddings for speaker recognition,''
\newblock in {\em ICASSP}, 2018, pp. 5329--5333.

\bibitem{okabe2018attentive}
Koji Okabe, Takafumi Koshinaka, and Koichi Shinoda,
\newblock ``Attentive statistics pooling for deep speaker embedding,''
\newblock in {\em INTERSPEECH}, 2018, pp. 2252--2256.

\bibitem{chung2018voxceleb2}
Joon~Son Chung, Arsha Nagrani, and Andrew Zisserman,
\newblock ``{VoxCeleb2}: Deep speaker recognition,''
\newblock in {\em INTERSPEECH}, 2018, pp. 1086--1090.

\bibitem{Jung2019raw}
Jee-weon Jung, Hee-Soo Heo, Ju-ho Kim, Hye-jin Shim, and Ha-Jin Yu,
\newblock ``{RawNet}: Advanced end-to-end deep neural network using raw
  waveforms for text-independent speaker verification,''
\newblock in {\em INTERSPEECH}, 2019, pp. 1268--1272.

\bibitem{Xie19a}
Weidi Xie, Arsha Nagrani, Joon~Son Chung, and Andrew Zisserman,
\newblock ``Utterance-level aggregation for speaker recognition in the wild,''
\newblock in {\em ICASSP}, 2019, pp. 5791--579.

\bibitem{Chen2019tied}
Nanxin Chen, Jes{\'u}s Villalba, and Najim Dehak,
\newblock ``Tied mixture of factor analyzers layer to combine frame level
  representations in neural speaker embeddings,''
\newblock in {\em INTERSPEECH}, 2019, pp. 2948--2952.

\bibitem{ding2018mtgan}
Wenhao Ding and Liang He,
\newblock ``{MTGAN}: Speaker verification through multitasking triplet
  generative adversarial networks,''
\newblock in {\em INTERSPEECH}, 2018, pp. 3633--3637.

\bibitem{Wang2019centroid}
Jixuan Wang, Kuan-Chieh Wang, Marc~T Law, Frank Rudzicz, and Michael Brudno,
\newblock ``Centroid-based deep metric learning for speaker recognition,''
\newblock in {\em ICASSP}, 2019, pp. 3652--3656.

\bibitem{bai2019partial}
Zhongxin Bai, Xiao-Lei Zhang, and Jingdong Chen,
\newblock ``Partial {AUC} optimization based deep speaker embeddings with
  class-center learning for text-independent speaker verification,''
\newblock in {\em ICASSP}, 2020, pp. 6819--6823.

\bibitem{Gao2019improving}
Zhifu Gao, Yan Song, Ian McLoughlin, Pengcheng Li, Yiheng Jiang, and Li-Rong
  Dai,
\newblock ``Improving aggregation and loss function for better embedding
  learning in end-to-end speaker verification system,''
\newblock in {\em INTERSPEECH}, 2019, pp. 361--365.

\bibitem{Sadjadi2019}
Seyed~Omid Sadjadi, Craig Greenberg, Elliot Singer, et~al.,
\newblock ``The 2018 {NIST} speaker recognition evaluation,''
\newblock in {\em INTERSPEECH}, 2019, pp. 1483--1487.

\bibitem{Ioffe06}
Sergey Ioffe,
\newblock ``Probabilistic linear discriminant analysis,''
\newblock in {\em ECCV}. Springer, 2006, pp. 531--542.

\bibitem{rahman2018improving}
Md~Hafizur Rahman, Ahilan Kanagasundaram, Ivan Himawan, David Dean, and Sridha
  Sridharan,
\newblock ``Improving {PLDA} speaker verification performance using domain
  mismatch compensation techniques,''
\newblock {\em Computer Speech \& Language}, vol. 47, pp. 240--258, 2018.

\bibitem{wang2018unsupervised}
Qing Wang, Wei Rao, Sining Sun, Lei Xie, Eng~Siong Chng, and Haizhou Li,
\newblock ``Unsupervised domain adaptation via domain adversarial training for
  speaker recognition,''
\newblock in {\em ICASSP}, 2018, pp. 4889--4893.

\bibitem{Williams2019ff}
Jennifer Williams and Simon King,
\newblock ``Disentangling style factors from speaker representations,''
\newblock in {\em INTERSPEECH}, 2019, pp. 3945--3949.

\bibitem{bhattacharya2019adapting}
Gautam Bhattacharya, Jahangir Alam, and Patrick Kenny,
\newblock ``Adapting end-to-end neural speaker verification to new languages
  and recording conditions with adversarial training,''
\newblock in {\em ICASSP}, 2019, pp. 6041--6045.

\bibitem{kang2020disentangled}
Woo~Hyun Kang, Sung~Hwan Mun, Min~Hyun Han, and Nam~Soo Kim,
\newblock ``Disentangled speaker and nuisance attribute embedding for robust
  speaker verification,''
\newblock {\em IEEE Access}, 2020.

\bibitem{shon2017autoencoder}
Suwon Shon, Seongkyu Mun, Wooil Kim, and Hanseok Ko,
\newblock ``Autoencoder based domain adaptation for speaker recognition under
  insufficient channel information,''
\newblock in {\em INTERSPEECH}, 2017, pp. 1014--1018.

\bibitem{zhang2018analysis}
Chunlei Zhang, Shivesh Ranjan, and John~HL Hansen,
\newblock ``An analysis of transfer learning for domain mismatched
  text-independent speaker verification,''
\newblock in {\em Proceedings of Odyssey: The Speaker and Language Recognition
  Workshop}, 2018, pp. 181--186.

\bibitem{lee2019coral}
Kong~Aik Lee, Qiongqiong Wang, and Takafumi Koshinaka,
\newblock ``The {CORAL+} algorithm for unsupervised domain adaptation of
  {PLDA},''
\newblock in {\em ICASSP}, 2019, pp. 5821--5825.

\bibitem{nagrani2017voxceleb}
Arsha Nagrani, Joon~Son Chung, and Andrew Zisserman,
\newblock ``{VoxCeleb}: a large-scale speaker identification dataset,''
\newblock {\em arXiv preprint arXiv:1706.08612}, 2017.

\bibitem{aishell_2017}
Hui Bu, Jiayu Du, Xingyu Na, Bengu Wu, and Hao Zheng,
\newblock ``Aishell-1: An open-source mandarin speech corpus and a speech
  recognition baseline,''
\newblock in {\em O-COCOSDA}, 2017, pp. 1--5.

\bibitem{wang2020remark}
Dong Wang,
\newblock ``Remarks on optimal scores for speaker recognition,''
\newblock CSLT@Tsinghua University, http://wangd.cslt.org/public/pdf/nl.pdf,
  2020.

\bibitem{kingma2014adam}
Diederik~P Kingma and Jimmy Ba,
\newblock ``{Adam}: A method for stochastic optimization,''
\newblock {\em arXiv preprint arXiv:1412.6980}, 2014.

\end{thebibliography}

\end{document}